\documentclass[letterpaper, 10 pt, conference]{ieeeconf}  

\usepackage{makeidx}
\usepackage{workpackages}

\usepackage[
backend=biber,
bibstyle=ieee,
citestyle=numeric-comp,
dashed=false,
url = false,
maxnames=2,
maxcitenames=2,
mincitenames=1,
sorting=none,
isbn = false,
doi = false]{biblatex}

\AtEveryBibitem{\clearlist{language}}
\AtEveryBibitem{\clearlist{location}}
\AtEveryBibitem{\clearfield{number}}
\AtEveryBibitem{\clearfield{month}}

\IEEEoverridecommandlockouts                              
\DeclareFieldFormat{labelnumberwidth}{[#1]\hspace{0.2em}}

\title{\LARGE \bf
Improving Cooperative Adaptive Cruise Control Robustness to Parametric Uncertainty via Plant Equivalent Controller Realizations}

    \author{Mischa Huisman, Thomas Arnold, Erjen Lefeber, Nathan van de Wouw, and Carlos Murguia 
\thanks{The research leading to these results has received funding from the European Union’s Horizon Europe programme under grant agreement No 101069748 – SELFY project.}
\thanks{M. Huisman, T. Arnold, C. Murguia, E. Lefeber, and N. van de Wouw are with the Department of Mechanical Engineering, Eindhoven University of Technology, The Netherlands. {\tt\small [M.R.Huisman}, {\tt\small C.G.Murguia}, {\tt\small A.A.J.Lefeber}, {\tt\small N.v.d.Wouw]@tue.nl}}
\thanks{C. Murguia is with the Department of Engineering Systems and Design, Singapore University of Technology and Design, Singapore. {\tt\small murguia\_rendon@sutd.edu.sg}}
}

\begin{document}

\maketitle
\thispagestyle{empty}
\pagestyle{empty}

\begin{abstract}
Cooperative Adaptive Cruise Control (CACC) enables vehicle platooning through inter-vehicle communication, improving traffic efficiency and safety. Conventional CACC relies on feedback linearization, assuming exact knowledge of vehicle parameters. However, the nonlinear longitudinal vehicle dynamics are subject to parametric uncertainty. Applying such feedback linearization with a nominal model yields imperfect cancellation, leading to model mismatch and degraded performance with off-the-shelf CACC controllers. To improve robustness without redesigning the CACC law, we explicitly model the mismatch between the ideal closed-loop dynamics, assumed by the CACC design, and the actual dynamics under parametric uncertainties. Robustness is formulated as an $\mathcal{L}_2$ trajectory-matching problem, minimizing the energy of this mismatch to make the uncertain system behave as closely as possible to the ideal model. This objective is addressed by optimizing over plant equivalent controller (PEC) realizations that preserve the nominal closed-loop behavior while mitigating the effects of parametric uncertainty. Stability and performance are enforced via LMIs, yielding a convex optimization problem applicable to heterogeneous platoons. Experimental results demonstrate improved robustness and performance under parametric uncertainty while preserving nominal behavior.

\end{abstract}

\section{INTRODUCTION} 
Cooperative Adaptive Cruise Control (CACC) is a well-explored technology for improving traffic throughput, fuel efficiency, and safety in intelligent transportation systems~\cite{guanetti_control_2018}. Many CACC designs, as well as other longitudinal controllers, rely on an additional nonlinear control layer that performs feedback linearization of the vehicle dynamics to facilitate subsequent linear controller design and analysis~\cite{ghasemi_stable_2013, ploeg_design_2011, lefeber_cooperative_2020, de_haan_cooperative_2025, wijnbergen_existence_2020}. However, feedback linearization inherently assumes exact knowledge of vehicle and road parameters. In practice, these parameters are uncertain and may vary significantly over time which can lead to instability and performance degradation.

To address model uncertainty in platoons, a variety of robust and adaptive control strategies have been proposed in the literature. Extensive research has focused on uncertainty in driveline dynamics, including online driveline parameter estimation~\cite{baldi_establishing_2021}, robust $\mathcal{H}_\infty$ control methods~\cite{gao_robust_2016,li_robust_2018}, and self-learning techniques~\cite{zhu_adaptive_2019}. Other approaches consider additional sources of uncertainty, such as wind gusts, which are addressed using robust model predictive control in~\cite{quan_robust_2024}. Nonlinear adaptive CACC schemes with online parameter updates have also been investigated~\cite{yang_adaptive_2023}, although such methods may be sensitive to measurement noise. Furthermore, adaptive control strategies that account for uncertain communication and actuation delays are proposed in \cite{xing_robust_2022, de_haan_cooperative_2025}.

Despite these advances, existing approaches typically focus on specific subsets of uncertainty or rely on simplified vehicle models, even though nonlinear dynamics and the associated uncertainties play an important role in, e.g., safety~\cite{alam_guaranteeing_2014}. Therefore, this paper first introduces a general nonlinear longitudinal vehicle model that explicitly accounts for multiple resistive forces, following established modeling approaches \cite{sheikholeslam_longitudinal_1993, sahlholm_road_2010, alam_guaranteeing_2014, besselink_cyberphysical_2016}. Feedback linearization is then applied without assuming exact knowledge of all model parameters.

To enhance closed-loop robustness against parametric uncertainty arising from imperfect cancellation in the feedback linearization, we exploit plant equivalent controller (PEC) realizations~\cite{huisman_optimal_2024,huisman_plant_2025}. By leveraging the degrees of freedom in the realization of a given dynamic CACC scheme, referred to herein as the \emph{base controller}, we mitigate the impact of such uncertainties while preserving the nominal closed-loop behavior. Specifically, by defining nominal and uncertainty-induced error states, the PEC realization can be designed to minimize transient and steady-state performance degradation.

The closed-loop system is formulated as a polytopic uncertain linear system, with the PEC realization design variables entering linearly. This structure enables robust stability and performance to be enforced through a set of LMIs evaluated at the vertices of the uncertainty polytope. The design objective combines the minimization of the average output energy induced by uncertainty with bounds on the worst--case input--output gain (i.e., $\mathcal{H}_\infty$). This yields a convex optimization problem that can be efficiently solved using SDP solvers. The proposed PEC realizations apply to heterogeneous platoons and explicitly optimize performance with respect to model uncertainty. The proposed method is experimentally validated, demonstrating improved robustness.

The remainder of this paper is organized as follows. Section~\ref{sec:Preliminaries} introduces the required preliminaries. Section~\ref{sec:ProblemSetting} presents the nonlinear vehicle model and introduces the PEC realizations. Section~\ref{sec:approach} describes the framework for robust stability and performance optimization using PEC realizations. Section~\ref{sec:Results} compares simulation and experimental results, and Section~\ref{sec:Conclusion} concludes the paper.

\section{Mathematical Preliminaries} \label{sec:Preliminaries}

The symbol $\mathbb{R}$ stands for the real numbers, $\mathbb{R}_{>0}$ ($\mathbb{R}_{\geq 0}$) denotes the set of positive (non-negative) real numbers. The symbol $\mathbb{N}$ denotes the set of natural numbers, including zero. The $n \! \times \! m$ matrix composed of only zeros is denoted by $0_{n \times m}$, or $0$ when its dimension is clear. The notation $A \succeq 0$ (resp., $A \preceq 0$) indicates that the matrix $A$ is positive (resp., negative) semidefinite, i.e., all the eigenvalues of the symmetric matrix $A$ are positive (resp., negative) or equal to zero, whereas $A \succ 0$ (resp., $(A \prec 0)$) indicates the positive (resp., negative) definiteness, i.e., all the eigenvalues are strictly positive (resp., negative). The signal 2-norm $\mathcal{L}_2$ is defined as $\|w\|_{\mathcal{L}_2} = \sqrt{\int_{-\infty}^{\infty} \|w(t)\|^2 dt}$. Time dependencies of signals are often omitted for simplicity of notation.


\begin{definition}[Polytopic Uncertain System] \
\label{def:polytopic_system}
Consider the linear system with model uncertainties~\cite{ding_model-based_2013,boyd_linear_1994}
\begin{subequations}\begin{align}
    \dot{x}(t) &= A(\Delta) x(t) + B(\Delta) u(t), \\
    y(t) &= C(\Delta) x(t) + D(\Delta) u(t),
\end{align}\end{subequations}
with state $x(t) \!\!\in\! \mathbb{R}^n$, input $u(t) \!\!\in\! \mathbb{R}^m$, output $y(t) \!\!\in\! \mathbb{R}^r$, and time $t$. For a polytopic uncertain system, matrices $A(\Delta), B(\Delta), C(\Delta),$ and $D(\Delta)$ are subject to parametric uncertainty $\Delta$ and belong to convex polytope $\Omega \! := \! \mathrm{conv}\!\left\{\!S^{[1]}\!, \dots, S^{[N]} \!\right\}$, with each vertex $S^{[k]}$ defined as tuple:
\begin{align}
    S^{[k]} :=0\left( A^{[k]}, B^{[k]}, C^{[k]}, D^{[k]} \right), \quad k = 1,\dots,N.
\end{align}
Accordingly, for any admissible uncertainty $\Delta$, the system matrices admit the representation
\begin{align}
    \Omega = \sum_{k=1}^{N} \alpha_k S^{[k]}, \, \sum_{k=1}^{N} \alpha_k = 1, \, \alpha_k \geq 0, \, k = 1,\dots,N. 
\end{align}
\end{definition}

\begin{lemma}[Quadratic Stability] \
\label{lemma1}
Consider the polytopic system described in Definition~\ref{def:polytopic_system} with $u(t) = 0$. If there exists a symmetric matrix $P \succ 0$ such that 
\begin{align}
    ({A}^{[k]})^\top P + P {A}^{[k]} \prec 0, \quad \forall k \in \{1, \dots, N\},
\end{align}
then the origin of the system is asymptotically stable for all admissible uncertainties $\Delta$~\cite{ding_model-based_2013,boyd_linear_1994}.
\end{lemma}

\begin{lemma}[Bounded Real Lemma] \
\label{lemma2}
Consider the polytopic system in Definition~\ref{def:polytopic_system} with performance output
$z(t) = C_z(\Delta) x(t)$. If there exists a symmetric matrix $P \succ 0$ and a scalar $\gamma > 0$ such that, for all vertices $k = 1,\dots,N$,
\begin{align}
    \begin{bmatrix}
        (A^{[k]})^\top P + P A^{[k]} + (C_z^{[k]})^\top C_z^{[k]} & P B^{[k]} \\
        (B^{[k]})^\top P & -\gamma I
    \end{bmatrix} \preceq 0,
\end{align}
then the $\mathcal{L}_2$-gain from $u$ to $z$ satisfies $\| z \|_{\mathcal{L}_2} \leq \sqrt{\gamma} \| u \|_{\mathcal{L}_2}$,
for all admissible uncertainties $\Delta$ and all $u \in \mathcal{L}_2$~\cite{ding_model-based_2013,boyd_linear_1994}.
\end{lemma}

\section{Problem Setting} \label{sec:ProblemSetting}
\begin{figure}[bt]
    \centering
    \includegraphics[width=\linewidth]{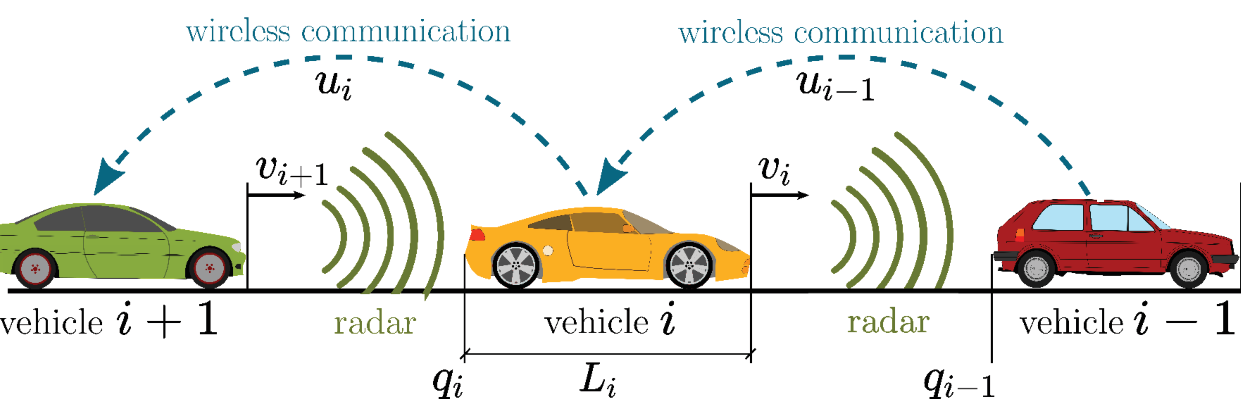}
    \caption{CACC-equipped string of vehicles.}
    \label{fig:SF_PlatoonModel}
\end{figure}

Consider a platoon of $m$ vehicles, schematically depicted in Fig. \ref{fig:SF_PlatoonModel}, indexed by $i\in\,S_m \coloneqq \{1,\ldots,m\}$, where $i=1$ denotes the lead vehicle. The nonlinear longitudinal dynamics of the $i$th vehicle are commonly modeled as~\cite{sheikholeslam_longitudinal_1993,sahlholm_road_2010,alam_guaranteeing_2014,besselink_cyberphysical_2016}
\begin{subequations} \label{eq:PF_NLModel} \begin{align}
    \dot{F}_{\text{drive},i} &= -F_{\text{drive},i}\tau_i^{-1} + \eta_i\tau_i^{-1}, \label{eq:PF_engDyn} \\
     m_{\text{eff},i} a_i &= F_{\text{drive},i} - C_{\text{d},i} (v_i-v_\text{w} )^2  \nonumber \\
     & \qquad - f_{\text{v},i} v_i - f_{\text{r},i} m_i g - m_i g \alpha_i, \label{eq:PF_accDyn}
\end{align}\end{subequations}
with effective vehicle mass $m_{\text{eff},i}$, static mass $m_i$, and driveline time constant $\tau_i$. The variables $F_{\text{drive},i}$ and $\eta_i$ represent the driving force and engine input, while $v_i$ and $a_i$ are the vehicle's velocity and acceleration, respectively. Environmental and resistive factors include the drag coefficient $C_{\text{d},i}$, wind speed $v_\text{w}$, viscous friction $f_{\text{v},i}$, rolling resistance $f_{\text{r},i}$, gravitational constant $g$, and road slope $\alpha_i$.

Assuming small road slopes $\alpha_i$, and slowly varying rolling and gravitational forces, differentiation of~\eqref{eq:PF_accDyn}, together with~\eqref{eq:PF_NLModel} as an expression for $F_{\text{drive},i}$,  yields the nonlinear acceleration dynamics
\begin{multline} \label{eq:PF_nlPlatoon}
     \hspace{-3.5mm}\dot{a}_i = - \tfrac{f_{\text{v},i} - 2C_{\text{d},i} v_\text{w}}{\tau_i m_{\text{eff},i}} v_i - \left(\tfrac{f_{\text{v},i} - 2C_{\text{d},i} v_\text{w}}{m_{\text{eff},i}} + \tfrac{1}{\tau_i} \right)a_i -\tfrac{C_{\text{d},i}}{\tau_i m_{\text{eff},i}}v_i^2 \\
    - \tfrac{2 C_{\text{d},i}}{m_{\text{eff},i}}v_i a_i - \tfrac{m_i(f_{\text{r},i} g + g \alpha_i) + C_{\text{d},i} v_\text{w}^2}{\tau_i m_{\text{eff},i}} + \tfrac{1}{\tau_i m_{\text{eff},i}} \eta_i.
\end{multline}
The control objective for each follower vehicle is to keep a desired inter-vehicle distance $d_{\text{r},i} = r_i + h_i v_i, \, i \in S_m \backslash \{ 1 \}$, with time gap $h_i>0$ and standstill distance $r_i\geq0$. Several CACC schemes have been proposed in the literature to achieve this objective, e.g.,~\cite{ploeg_design_2011,lefeber_cooperative_2020, de_haan_cooperative_2025}, which are based on a vehicle model obtained via feedback linearization of~\eqref{eq:PF_nlPlatoon}.

Feedback linearization is achieved via engine input $\eta_i$, and requires exact knowledge of all parameters in~\eqref{eq:PF_nlPlatoon}. Since this assumption is unrealistic, we introduce six parameter groups $p_j$, $j \in \{1,\ldots,6\}$, associated with the six terms in~\eqref{eq:PF_nlPlatoon}, and each characterized by a nominal estimate $p_{j|0}$ and an additive uncertainty $\Delta_j$. This allows rewriting~\eqref{eq:PF_nlPlatoon} as
\begin{multline}\label{eq:PF_nl_Uncertain}
   \! \! \! \dot{a}_i =  -(p_{1|0} + \Delta_1)v_i 
    - (p_{2|0} + \Delta_2)a_i 
    - (p_{3|0} + \Delta_{3})v_i^2 \\ 
    - (p_{4|0} + \Delta_{4}) v_i a_i 
    - (p_{5|0} + \Delta_5) 
    + (p_{6|0} + \Delta_6) \eta_i
\end{multline}
with uncertainty vector $\Delta$ and hyper-rectangle set $\mathcal{D}$
\begin{subequations}\begin{align}
    \Delta &\coloneqq 
    \begin{bmatrix}
        \Delta_1 & \Delta_2 & \Delta_3 & \Delta_4 & \Delta_5 & \Delta_6
    \end{bmatrix}^\top, \\
    \label{eq:uncertainConvexHull}
    \hspace{-1mm} \Delta \! \in \! \mathcal{D} &:= \left\{ \Delta \! \in \! \mathbb{R}^6 : \Delta_j \in [\underline{\Delta}_j, \overline{\Delta}_j], \; j \in \{1,\dots,6\} \right\}.
\end{align}\end{subequations}

Following~\cite{ploeg_design_2011,lefeber_cooperative_2020}, the desired vehicle model is defined as
\begin{align} \label{eq:desiredModel}
    \dot{q}_i &= v_i, \qquad \dot{v}_i = a_i, \qquad
    \dot{a}_i = -\tau_{\text{des},i}^{-1} a_i + \tau_{\text{des},i}^{-1} u_i, 
\end{align}
with position $q_i$, velocity $v_i$, new input $u_i$, and design parameter $\tau_{\text{des},i} > 0$. Using the nominal parameter estimates $p_{j|0}$, the feedback linearizing control law is given by
\begin{multline} \label{eq:PF_IOcontroller_param}
    \eta_i = p_{6|0}^{-1} \Big(
        p_{1|0} v_i
        + \Big(p_{2|0} - \tau_{\text{des},i}^{-1} \Big) a_i
        + p_{3|0} v_i^2  \\
        + p_{4|0} v_i a_i
        + p_{5|0}
        + \tau_{\text{des},i}^{-1} u_i
    \Big).
\end{multline}
Substituting~\eqref{eq:PF_IOcontroller_param} into~\eqref{eq:PF_nl_Uncertain} yields the perturbed dynamics
\begin{subequations}\label{eq:PF_simplified_uncertain}
\begin{align}
    \dot{a}_i &= -\tau_{\text{des},i}^{-1} a_i + \tau_{\text{des},i}^{-1} u_i + \phi_i, \\
    \phi_i &= \theta_{i,1} v_i
    + \theta_{i,2} a_i  \nonumber \\
    & \qquad + \theta_{i,3} v_i^2 
    + \theta_{i,4} v_i a_i
    + \theta_{i,5} 
    + \theta_{i,6} u_i, \\
    \theta_{i,1} &= \tfrac{p_{1|0}}{p_{6|0}}\Delta_6 - \Delta_1, \label{eq:theta1} \\
    \theta_{i,2} &= \left(\tfrac{p_{2|0}}{p_{6|0}} - \tfrac{1} {p_{6|0}\tau_{\text{des},i}} \right) \Delta_6 - \Delta_2, \\
    \theta_{i,3} &= \tfrac{p_{3|0}}{p_{6|0}}\Delta_6 - \Delta_3, \qquad
    \theta_{i,4} = \tfrac{p_{4|0}}{p_{6|0}} \Delta_6 - \Delta_4, \\
    \theta_{i,5} &= \tfrac{p_{5|0}}{p_{6|0}} \Delta_6 - \Delta_5, \qquad 
    \theta_{i,6} = \tfrac{1}{p_{6|0}\tau_{\text{des},i}} \Delta_6. \label{eq:theta6}
\end{align}
\end{subequations}
In the absence of uncertainty (i.e., $\Delta = 0 \!\!\implies\!\! \phi_i=0$), the perturbed model~\eqref{eq:PF_simplified_uncertain} reduces to the desired model~\eqref{eq:desiredModel}. The CACC schemes for heterogeneous platoons~\cite{ploeg_design_2011,lefeber_cooperative_2020} are typically designed using the nominal model~\eqref{eq:desiredModel}. Consequently, when applied to real vehicles where $\phi_i \neq 0$, the desired nominal performance is not achieved.

To make this explicit, consider the CACC scheme for heterogeneous platoons~\cite{ploeg_design_2011,lefeber_cooperative_2020}:
\begin{align}
    \label{eq:SD_BaseController} 
    \mathcal{F} &\coloneqq \left\{
    \begin{aligned}
        \dot{\rho}_i &= -\tfrac{1}{h} \rho_i + \tfrac{k_p}{h_i}e_i + \tfrac{k_d}{h_i} \dot{e}_i \\ & \qquad + \tfrac{\tau\imo - \tau_i}{h_i \tau\imo}\aim + \tfrac{\tau_i}{h_i \tau\imo}\uim,\\[-1ex]
        u_i &= \rho_i, 
    \end{aligned} \right.          
\end{align}
with spacing error $e_i \coloneqq d_i - d_{\text{r},i}$, inter-vehicle distance $d_i = q_{i-1} - q_i - L_i$, vehicle length $L_i$, and controller gains ($k_p,k_d$). Let $x_i = [d_i\, v_i\, a_i]^\top$ and $x\imo =[\vim \, \aim]^\top$ denote the follower and leader state, respectively. Assuming a perfect leader (i.e., $\phi_{i-1}=0$), by substituting~\eqref{eq:SD_BaseController} into~\eqref{eq:PF_simplified_uncertain} we obtain the perturbed closed-loop system
\begin{align} \label{eq:CL_base_uncertain}
     & \hspace{-3mm} \begin{bmatrix}
        \dot{x}\imo\\ \dot x_i \\ \dot{{\rho}}_i
    \end{bmatrix} \!\!=\!\! 
    \begin{bmatrix}
        A^0_{i-1,i-1} & 0 & 0\\ 
        A^0_{i,i-1} & A^0_{i,i} &  B^0_{i,i} \\
        {B}^c_{i,i-1} & {B}^c_{i,i} & {A}^c_i 
    \end{bmatrix}\!\!
    \begin{bmatrix}
        x\imo \\ x_i \\ {\rho}_i
    \end{bmatrix} \nonumber \\ 
    & \hspace{-2mm} + \! \begin{bmatrix}
        (B^0_{i-1})^\top & 
        0     &
        (E^c_i)^\top     
    \end{bmatrix}^\top  
        \!\! \uim \! + \! 
    \begin{bmatrix}
        0 & 
        (B^\phi_i)^\top&
        0     
    \end{bmatrix}^\top  
        \!\! \phi_i,
\end{align}
where superscripts $0$ and $c$ denote nominal and controller matrices, respectively, given in Appendix~\ref{app:appendix1}. While the base controller~\eqref{eq:SD_BaseController} achieves the desired nominal performance for~\eqref{eq:desiredModel}, the presence of $\phi_i$ in the true closed-loop system~\eqref{eq:CL_base_uncertain} leads to performance degradation.

Rather than designing an entirely new robust controller, we enhance the robustness of~\eqref{eq:CL_base_uncertain} while preserving nominal closed-loop behavior. While the base controller~\eqref{eq:SD_BaseController} guarantees nominal performance by design, its internal realization is not unique. This non-uniqueness is exploited through PEC realizations~\cite{huisman_optimal_2024,huisman_plant_2025}, i.e., distinct controller realizations that produce identical control inputs $u_i$ in the nominal case but possess different internal dynamics that can be tuned to mitigate the impact of uncertainties.

The PEC realization of~\eqref{eq:SD_BaseController} is obtained by defining a new controller state $\bar{\rho}_i$ via the linear transformation:
\begin{align} \label{eq:PEC_Transformation}
  \!\!\! \bar{\rho}_i =  \rho_i + F_{i,i} x_i + F_{i,i-1} x\imo,
\end{align} 
where $F_{i,i} \in \mathbb{R}^{1 \times 3}$ and $F_{i,i-1} \in \mathbb{R}^{1 \times 2}$ are design variables. Differentiating~\eqref{eq:PEC_Transformation} and substituting the nominal dynamics~\eqref{eq:desiredModel} and~\eqref{eq:SD_BaseController} yields the PEC realization
\begin{subequations}\label{eq:PEC_Controller} \begin{align}  \hspace{-3mm}
    \bar{\mathcal{F}} &\coloneqq \left\{
    \begin{aligned}
        \dot{\bar{\rho}}_i &= \bar{A}^c_i \bar{\rho}_i + \bar{B}^c_{i,i} x_i + \bar{B}^c_{i,i-1} x\imo + \bar{E}^c_i \uim, \\
        u_i &= \bar{\rho}_i - F_{i,i} x_i - F_{i,i-1} x\imo,
    \end{aligned} \right.\\
    & \hspace{-2mm} \bar{A}^c_i = A_i^c + F_{i,i} B^0_{i}, \quad {\bar{E}}^c_i = E^c_i + F_{i,i-1} B^0_{i-1}, \\
    & \hspace{-2mm} \bar B^c_{i,i} = B^c_{i,i} + F_{i,i} A^0_{i,i} - F_{i,i} B^0_{i,i} F_{i,i}- A^c_i F_{i,i} \\
    & \hspace{-2mm} {\bar{B}}^c_{i,i-1} = {B}^c_{i,i-1} + F_{i,i-1} A^0_{i-1,i-1} + F_{i,i} A^0_{i,i-1} \nonumber \\
    & \qquad - F_{i,i} B^0_{i,i} F_{i,i-1} - A^c_i F_{i,i-1}. 
\end{align}
\end{subequations}
 
The problem addressed in this paper is to determine a PEC realization that improves the robustness of a CACC-controlled vehicle in the presence of parametric uncertainty. Specifically, given the base controller~\eqref{eq:SD_BaseController}, we seek to design the matrices $F_{i,i}$ and $F_{i,i-1}$ in~\eqref{eq:PEC_Controller} such that the influence of the uncertainty $\phi_i$ on the closed-loop system is minimized, while preserving the nominal tracking performance.


\section{Optimal PEC Design} \label{sec:approach} 
This section derives the closed-loop dynamics under the PEC realization while accounting for parametric uncertainty. Without loss of generality, we restrict the problem to a two-vehicle platoon, i.e., $S_m = \{1,2\}$, and focus on the follower vehicle $i=2$. This allows us to avoid cumbersome subscript notation in the remainder of the paper while preserving the essence of the problem. Accordingly, we drop the subscript $i$ from $F_{i,i}$ and $F_{i,i-1}$ and define $F \coloneqq[F_{2} \,\, F_{1}] = [f_{2,1}, f_{2,2}, f_{2,3}, f_{1,1}, f_{1,2}]$. After deriving the closed-loop dynamics, we formulate an optimization problem to derive the PEC realization that minimizes the impact of the parametric uncertainty $\phi_2$ on the closed-loop behavior.

\subsection{Uncertain Closed-Loop Dynamics} \label{sec:uncertainCLdyn}  

Substituting the PEC realization~\eqref{eq:PEC_Controller} into the uncertain dynamics~\eqref{eq:PF_simplified_uncertain} yields the perturbed closed-loop system
\begin{multline} \label{eq:CL_PEC_uncertain} 
    \hspace{-4mm}\begin{bmatrix}
        \dot{x}_1\\ \dot x_2 \\ \dot{\bar{\rho}}_2
    \end{bmatrix} \!\!=\!\! 
    \begin{bmatrix}
        A^0_{1,1} & 0 & 0\\ 
        A^0_{2,1} - B^0_{2} F_{1} & A^0_{2,2} - B^0_{2} F_{2} &  B^0_{2} \\
        \bar{B}^c_{2,1} & \bar{B}^c_{2,2} & \bar{A}^c_2 
    \end{bmatrix}\!\!
    \begin{bmatrix}
        x_1 \\ x_2 \\ \bar{\rho}_2
    \end{bmatrix} \\ +  
    \begin{bmatrix}
        (B^0_1)^\top  &
        0     &
        \bar{E}^c_2     
    \end{bmatrix}^\top  
        u_1 + 
    \begin{bmatrix}
        0 & 
        (B^\phi_2)^\top   &
        0     
    \end{bmatrix}^\top   
        \phi_2,
\end{multline}
with the system matrices defined in Appendix~\ref{app:appendix1}. By construction, the nominal closed-loop behavior of~\eqref{eq:CL_PEC_uncertain} is equivalent to that of~\eqref{eq:CL_base_uncertain} when $\phi_2 = 0$. However, the choice of PEC realization matrix $F$ affects the closed-loop behavior in the presence of parametric uncertainty, i.e., when $\phi_2 \neq 0$.

To explicitly characterize the influence of the  PEC realization, we rewrite the closed-loop dynamics~\eqref{eq:CL_PEC_uncertain} in the original controller coordinates $\rho_2$ by applying the inverse transformation of~\eqref{eq:PEC_Transformation}. In the case of uncertainties, this yields
\begin{align} \label{eq:CL_base_F}
    \hspace{-2mm}\begin{bmatrix}
        \dot{x}_1\\ \dot x_2 \\ \dot{{\rho}}_2
    \end{bmatrix} \!\! &=\!\! 
    \begin{bmatrix}
        A^0_{1,1} & 0 & 0\\ 
        A^0_{2,1} & A^0_{2,2} &  B^0_{2,2} \\
        {B}^c_{2,1} & {B}^c_{2,2} & {A}^c_2 
    \end{bmatrix}\!\!
    \begin{bmatrix}
        x_1 \\ x_2 \\ \rho_2
    \end{bmatrix}  \! +  \!
    \begin{bmatrix}
        B^0_1 \\
        0    \\
        E^c_2     
    \end{bmatrix}  
       u_1 \nonumber \\
        & \qquad + 
    \begin{bmatrix}
        0 &
        (B^\phi_2)^\top  &
        -F_{2} B^\phi_2     
    \end{bmatrix}^\top  
        \phi_2.
\end{align}
From~\eqref{eq:CL_base_F}, it is evident that the PEC realizations only affect the mapping of the uncertainty $\phi_2$ into the closed-loop dynamics. Moreover, this dependence occurs through the product $F_2 B^\phi_2$, which reduces to the scalar parameter $f_{2,3}$. Consequently, although $F \!\in \!\mathbb{R}^{1 \times 5}$, the effect of the PEC realization with respect to parametric uncertainty is fully characterized by the parameter $f_{2,3}$.

Since this result holds without imposing any assumptions on the structure of $\phi_2$, the remaining PEC parameters can be fixed without affecting robustness. In particular, inspired by~\cite{lefeber_cooperative_2020}, we choose $f_{1,2} = -\tfrac{\tau_2}{h_2}$, which yields PEC realizations that are independent of the leader input $u_1$ (i.e., $\bar{E}^c_2 = 0$) and driveline time constant $\tau_{1}$. The latter property is especially relevant for heterogeneous platoons, as it avoids dependency on the driveline dynamics of the preceding vehicle.

To determine the optimal design for $f_{2,3}$, the nonlinear uncertainty $\phi_2$ is approximated by a first-order Taylor expansion around an operating point $(v_2^*, a_2^*, u_2^*)$. Assuming both the leader and follower vehicles travel at a constant reference velocity $v_{\mathrm{r}}$, we set the operating point to $(v_2^*, a_2^*, u_2^*) = (v_{\mathrm{r}}, 0, 0)$. Under this assumption, the closed-loop system can be approximated by the following linear, uncertain system:
\begin{subequations} \label{eq:CL_uncertain2} 
\begin{align} 
    & \begin{bmatrix}
        \dot x_1 \\ \dot x_2 \\ \dot{{\rho}}_2
    \end{bmatrix} = \left(
    \begin{bmatrix}
        A^0_{1,1} & 0 & 0 \\ 
        A^0_{2,1} & A^0_{2,2} &  B^0_{2,2} \\
        {B}^c_{2,1} & {B}^c_{2,2} & {A}^c_2 
    \end{bmatrix} \right. \nonumber \\ & \left. + 
    \begin{bmatrix}
        0 & 0 & 0\\ 
        0 & A^\Delta_{2} & B^\Delta_{2} \\
        0 & -F_{2} A^\Delta_{2} & -F_{2} B^\Delta_{2} 
    \end{bmatrix} \right)
    \begin{bmatrix}
        x_1 \\ x_2 \\ \rho_2
    \end{bmatrix}  + 
    \begin{bmatrix}
        B^0_1 \\ 0 \\ 0
    \end{bmatrix} u_1 \nonumber \\ 
    & \,+ 
    \begin{bmatrix}
        0 & (B^\Delta_{0})^\top & -F_{2}B^\Delta_{0}
    \end{bmatrix}^\top, \\
    \hspace{-2mm}A^\Delta_2 \hspace{-1mm}&=\hspace{-1mm} \sum_{j=1}^6 \! A^\Delta_{2,j} \Delta_j, \, B^\Delta_2 \hspace{-1mm}=\hspace{-1mm} \sum_{j=1}^6 \! B^\Delta_{2,j} \Delta_j, \,
    B^\Delta_0 \hspace{-1mm} = \hspace{-1mm} \sum_{j=1}^6 \!B^\Delta_{0,j} \Delta_j. 
 \end{align}
\end{subequations} 
The matrices $A^\Delta_{2,j}, B^\Delta_{2,j}, B^\Delta_{0,j}$ are detailed in Appendix~\ref{app:uncertain}.

Since $A^\Delta_{2}, B^\Delta_{2}$, and $B^\Delta_{0}$ depend linearly on the uncertainty vector $\Delta$, their image over the set $\mathcal{D}$ (defined in~\eqref{eq:uncertainConvexHull}) lies within the convex hull of the matrices evaluated at the vertices of $\mathcal{D}$. There are $2^6$ vertices, corresponding to all binary combinations of lower and upper bounds. For each vertex $k\in\{1,2,\dots,2^6\}$, let $v^{[k]} \in \{0,1\}^6$ indicate whether to use the lower ($v^{[k]}_j = 0$) or upper ($v^{[k]}_j = 1$) bound for the $j$-th parameter. Then, for any $\Delta \in \mathcal{D}$, we have:
\begin{subequations}\begin{align}
    \Delta^{[k]}_j &= (1-v^{[k]}_j)\,\underline{\Delta}_j + v^{[k]}_j \,\overline{\Delta}_j, \\
     A^{\Delta|[k]}_{2} &= \sum_{j=1}^6 A^\Delta_{2,j} \Delta^{[k]}_j,  \quad
    B^{\Delta|[k]}_2 = \sum_{j=1}^6 B^\Delta_{2,j} \Delta^{[k]}_j, \\
    B^{\Delta|[k]}_0 &= \sum_{j=1}^6 B^\Delta_{0,j} \Delta^{[k]}_j, \qquad  \forall k\in \{1,...,2^6\}.
\end{align}\end{subequations}
Consequently, the uncertain matrices belong to the polytope:
\begin{subequations}\label{eq:uncertainPolytopic}\begin{align}
    \hspace{-2mm} A^\Delta_2 \!&=\! \sum_{k=1}^{2^6}\sum_{j=1}^6 \beta_k  A^\Delta_{2,j} \Delta^{[k]}_j, 
    \,\, B_2^\Delta \!=\! \sum_{k=1}^{2^6}\sum_{j=1}^6 \beta_k B^\Delta_{2,j} \Delta^{[k]}_j, \\
    \hspace{-2mm} B^\Delta_0 \!&=\! \sum_{k=1}^{2^6}\sum_{j=1}^6 \beta_k  B^\Delta_{0,j} \Delta^{[k]}_j,     
    \quad \sum_{k=1}^{2^6} \beta_k \!= \!1, \quad \beta_k \geq 0. 
\end{align}\end{subequations}

The closed-loop system~\eqref{eq:CL_uncertain2} combined with the polytopic representation~\eqref{eq:uncertainPolytopic} constitutes a polytopic uncertain system (Definition~\ref{def:polytopic_system}), enabling the use of Lemmas~\ref{lemma1} and~\ref{lemma2} to compute the optimal realization $f_{2,3}$.

\subsection{Optimal PEC Realization} \label{Optimal PEC Realization}

The objective is to design the optimal PEC realization by determining the value of $f_{2,3}$ (embedded in $F_2$) that minimizes the impact of the parametric uncertainty $\Delta$ on the closed-loop behavior. Specifically, we seek to minimize the deviation of the uncertain closed-loop system from the desired nominal performance. To this end, we treat the uncontrollable leader state $x_1$ as an exogenous input and restrict the stability and performance analysis to the follower vehicle. Assuming the leader tracks a constant reference velocity $v_{\mathrm{r}}$, we define the leader error state $\tilde{x}_1 \coloneqq [\tilde{v}_1, a_1]^\top$, where $\tilde{v}_1 = v_1 - v_{\mathrm{r}}$, and assume that $\tilde{x}_1  \!\in\! \mathcal{L}_2$.

To quantify the deviation induced by the uncertainty, we first define the reduced closed-loop state $\zeta \coloneqq \begin{bmatrix} x_2^\top & \rho_2  \end{bmatrix}^\top$ and rewrite the dynamics~\eqref{eq:CL_uncertain2} as 
\begin{subequations} \label{eq:zeta_uncertain}
    \begin{align}
    \hspace{-3mm} \dot\zeta &= (\mathcal{A}^0 + \mathcal{A}^\Delta) \zeta + \mathcal{B}^0_1 \tilde{x}_1+ \mathcal{B}^\Delta_0, \\
    \hspace{-3mm} \mathcal{A}^0 &=     \begin{bmatrix}
        A^0_{2,2} &  B^0_{2} \\
        {B}^c_{2} & {A}^c_2 
    \end{bmatrix}, \quad \mathcal{A}^\Delta =    
    \begin{bmatrix}
        A^\Delta_{2} & B^\Delta_{2} \\
        -F_{2} A^\Delta_{2} & -F_{2} B^\Delta_{2} 
    \end{bmatrix}, \\
    \hspace{-3mm} \mathcal{B}^0_1 \!&=\! 
    \begin{bmatrix}
        (A^0_{2,1})^\top & {B}^c_{2,1} 
    \end{bmatrix}^\top, 
    \mathcal{B}^\Delta_0 \!=\!
    \begin{bmatrix}
        (B^\Delta_{0})^\top & -F_{2}B^\Delta_{0}
    \end{bmatrix}^\top\!\!\!.
\end{align}
\end{subequations}
Then, let the nominal model (where $\Delta=0$) be defined as:
\begin{align}\label{eq:zeta_nominal}
    \dot\zeta^0 = \mathcal{A}^0 \zeta^0 + \mathcal{B}_1^0 \tilde x_1.
\end{align}

By defining the closed-loop error state $\zeta^e \coloneqq \zeta - \zeta^0$ and utilizing~\eqref{eq:zeta_uncertain} and~\eqref{eq:zeta_nominal}, the error dynamics yield:
\begin{align} \label{eq:errorDyn}
    \dot\zeta^e = (\mathcal{A}^0 + \mathcal{A}^\Delta)\zeta^e
      + \mathcal{A}^\Delta \zeta^0
      + \mathcal{B}_0^\Delta.
\end{align}
Based on~\eqref{eq:errorDyn}, the design objective is twofold: 1) reducing the coupling between the nominal state $\zeta^0$ and the error state $\zeta^e$, which characterizes how the nominal trajectory perturbs the actual dynamics; 2) minimizing the impact of the leader state $\tilde{x}_1$ on the error dynamics.

To achieve these objectives via Lemma~\ref{lemma1} (Quadratic Stability) and Lemma~\ref{lemma2} (Bounded Real Lemma), consider augmented state $q :=[ {\zeta^0}^\top  {\zeta^e}^\top]^\top$ and performance output $z$:
    \begin{subequations} \label{eq:qDyn_all} \begin{align} \label{eq:qDyn}
    \dot q &=
    \underbrace{\begin{bmatrix}
        \mathcal{A}^0 & 0 \\
        \mathcal{A}^\Delta & \mathcal{A}^0 + \mathcal{A}^\Delta
    \end{bmatrix}}_{\mathcal{A}_q} q
    +
   \underbrace{\begin{bmatrix}
        \mathcal{B}_1^0 \\ 0
    \end{bmatrix}}_{\mathcal{B}_q} \tilde x_1
    +
    \begin{bmatrix}
        0 \\ \mathcal{B}_0^\Delta
    \end{bmatrix}, \\
    z &\coloneqq  
    \begin{bmatrix}
        0 & C
    \end{bmatrix} q \eqqcolon C_z q, \label{eq:perfomanceOutput} \quad 
    C = \mathrm{diag}([1,1,1,0]).
\end{align}\end{subequations}
Since $\mathcal{A}_q$ depends linearly on $F_2$,~\eqref{eq:qDyn_all} provides a suitable framework for LMI-based optimization. The performance output $z$ targets only the vehicle-related error states within $\zeta^e$, as we are solely interested in improving the vehicle's behavior rather than the controller state $\rho_2$.

To achieve objective 1, we minimize $\mathrm{trace}(C_z P C_z^\top)$ to bound the average performance output energy. To achieve objective 2, we minimize the $\mathcal{L}_2$-gain $\gamma$ from the leader state $\tilde{x}_1$ to the performance output $z$~\cite{boyd_linear_1994}. Combining these two objectives and applying Lemma~\ref{lemma1} and Lemma~\ref{lemma2}, for a given $f_{2,3}$, if there exist a positive definite matrix $P \in \mathbb{R}^{8\times8}$ and a scalar $\gamma>0$, being the solution to the convex program: 
\begin{align} \label{eq:OP_OptProblem}
	\left\{
	\begin{aligned}
        &\underset{P, \gamma, f_{2,3}}{\min}
        && \mathrm{trace}(C_z P C_z^\top) + \gamma,\\
		&\text{\emph{s.t.}}
        && (\mathcal{A}_q)^\top P + P\mathcal{A}_q \prec 0,
        \quad k = 1,\ldots,2^6,\\
        &&&
        \begin{bmatrix}
			(\mathcal{A}_q)^\top P + P\mathcal{A}_q + C_z^\top C_z & P\mathcal{B}_q \\
			(\mathcal{B}_q)^\top P & -\gamma I
		\end{bmatrix}
        \preceq 0,
	\end{aligned}
	\right.
\end{align}
then the optimal value $f_{2,3}^*$ ensures robust stability and performance. The multi-objective formulation reflects a trade-off between reducing error dynamic energy and suppressing leader-induced deviations. Since \eqref{eq:OP_OptProblem} is bilinear in $P$ and $f_{2,3}$, a line-search over $f_{2,3}$ is used to maintain an LMI.

The proposed optimization framework is experimentally validated in the subsequent section.


\section{Case Study Results} \label{sec:Results}
This section presents the design of the optimal PEC realization obtained by solving~\eqref{eq:OP_OptProblem}, followed by an evaluation on both MATLAB simulations and experiments on a Renault Twizy test platform. For both scenarios, we consider a two-vehicle platoon, with leader $i=1$ and follower $i=2$; in the experimental setup, the Renault Twizy tracks a virtual leader. The vehicle parameters and the associated parametric uncertainties (considered only for the follower vehicle) are summarized in Table~\ref{tab:vehicle_params}. The problem is solved in MATLAB~2025b using YALMIP~\cite{lofberg_yalmip_2004}, with solver MOSEK~\cite{mosek_aps_mosek_2024}.
\begin{table}[bt]
\centering
\footnotesize
\setlength{\tabcolsep}{3pt}
\caption{Vehicle Parameters and estimated uncertainties.}
\label{tab:vehicle_params} 
\begin{tabular}{llccc}
\toprule
\textbf{Sym.} & \textbf{Parameter} & \textbf{Nominal} & \textbf{Range ($\Delta$)} & \textbf{Unit} \\ 
\midrule
$m$              & Vehicle Mass          & 731      & $[716, 746]$       & kg \\
$m_{\text{eff}}$ & Effective Mass        & 778      & $[763, 793]$       & kg \\
$f_v$            & Viscous Friction      & 5.55     & $[5.5, 5.6]$       & Ns/m \\
$f_r$            & Rolling Res. Coeff.   & 0.0262   & $[0.0262, 0.0262]$ & -- \\
$C_d$            & Aero. Drag Factor     & 0.392    & $[0.3528, 0.4312]$ & kg/m \\
$\tau$           & Time Constant         & 0.12     & $[0.11, 0.13]$     & s \\
$\alpha$         & Road Slope            & 0        & $[0, 0]$           & rad \\
$v_{\text{w}}$   & Wind Velocity         & 0        & $[-5, 5]$          & m/s \\
$(k_p,k_d)$      & Prop./Deriv. gain     & (0.2,0.7)& --                 & (1/s$^2$,1/s) \\ 
$h$              & Headway time          & 0.2      & --                 & s \\ 
\bottomrule
\end{tabular}  
\end{table} 

Five different PEC realizations, denoted by $F^l,\,l\in\{0,\ldots,4\}$, are evaluated. $F^0$ is the base controller. $F^1$ modifies all entries except for $f_{2,3}$, which, as discussed in Section~\ref{sec:uncertainCLdyn}, theoretically yields identical closed-loop behavior to $F^0$ under uncertainty. Realizations $F^2$, $F^3$, and $F^4$ are obtained by solving~\eqref{eq:OP_OptProblem} with different objective functions; while~\eqref{eq:OP_OptProblem} jointly minimizes $\mathrm{trace}(C_z P C_z^\top)$ and $\gamma$, the optimization problem is solved for each objective individually to isolate their respective effects. The optimization problem is solved for a range of operating points $v_\text{r} \in [0,100]$~km/h. Although the optimal cost increases with $v_\text{r}$, the optimizer $f_{2,3}^*$ appeared invariant. Consequently, all subsequent results are obtained using $v_\text{r} = 15 $~km/h. The resulting PEC matrices are summarized in TABLE~\ref{tab:realizations}. 
\begin{table}[bt]
\centering
\footnotesize
\setlength{\tabcolsep}{3pt}
\caption{PEC Realizations for Case Study Results.}
\label{tab:realizations}
\begin{tabular}{llll}
\toprule
 & \textbf{Objective} & $\mathbf{F}$ \\ 
\midrule
$F_0$ & Base controller & $[0,\,0,\,0,\, 0,\,-\tfrac{\tau_i}{h_i}]$ \\
$F_1$ & Equal performance & $[-\tfrac{\tau_i}{h_i},\,-\tfrac{\tau_i}{h_i},\,0,\,-\tfrac{\tau_i}{h_i},\,-\tfrac{\tau_i}{h_i}]$ \\
$F_2$ & Minimal $\mathrm{trace}(C_z P C_z^\top)$ + $\gamma$ & $[0,\,0,\,0.14865,\,0,\,-\tfrac{\tau_i}{h_i}]$ \\
$F_3$ & Minimal $\mathrm{trace}(C_z P C_z^\top)$ & $[0,\,0,\,0.15866,\,0,\,-\tfrac{\tau_i}{h_i}]$ \\
$F_4$ & Minimal $\gamma$ & $[0,\,0,\,0.11662,\,0,\,-\tfrac{\tau_i}{h_i}]$ \\
\bottomrule
\end{tabular} 
\end{table} 

\subsection{Simulation Results} 
In the simulation, we evaluate the nonlinear model~\eqref{eq:CL_PEC_uncertain} with a constant wind speed $v_\text{w} = 15$~km/h, and set the follower vehicle parameters at their lower bounds (TABLE~\ref{tab:vehicle_params}) to introduce a model mismatch. The leader tracks a reference velocity $v_\text{r}=15$~km/h and applies the step input $u_1(t) = 1$ for $t \in [10, 11]$, $-1$ for $t \in [21, 22]$, and $0$ otherwise.

\begin{figure}[bt]
    \centering
    \includegraphics[width=\linewidth]{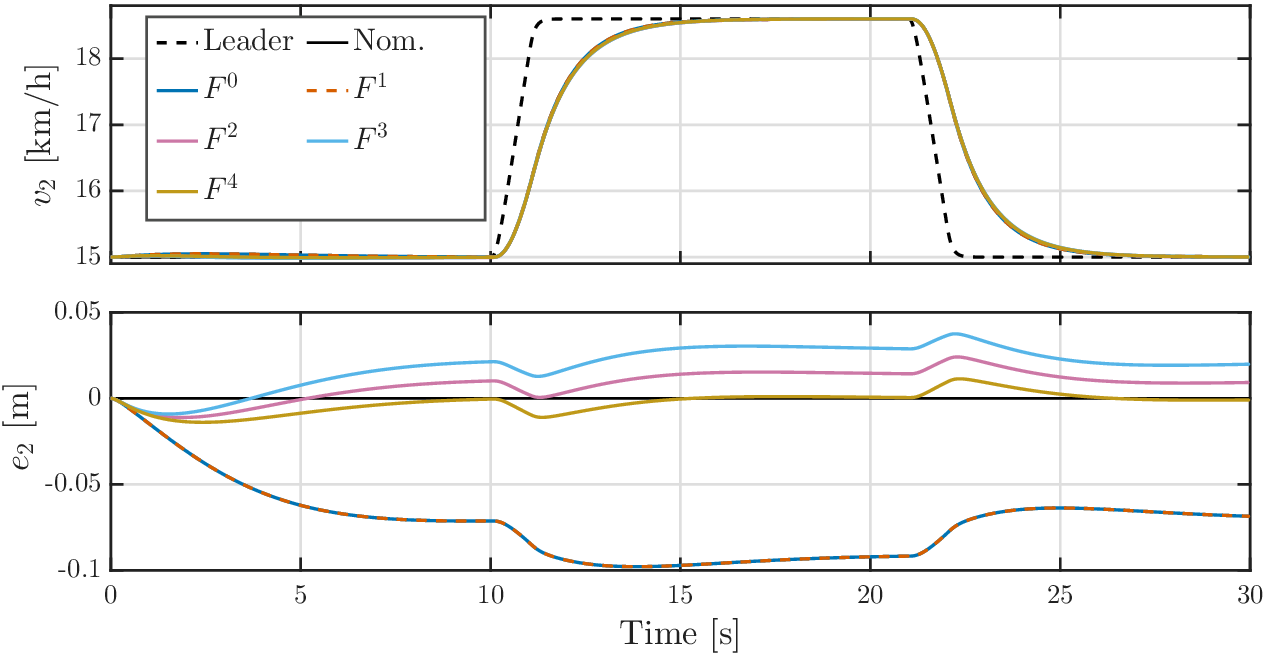} 
    \caption{Follower velocity $v_2$ and spacing error $e_2$ for the PEC realizations $F^l, l\in\{1,...,4\}$, in TABLE~\ref{tab:realizations}.}
    \label{fig:sim_result}
\end{figure}

The resulting follower velocity and spacing error for each realization $F^l, \, l \in \{1, \dots, 4\}$ are illustrated in Fig.~\ref{fig:sim_result}, with the nominal response ($\Delta = 0$) provided as a baseline (solid black line). To quantify the deviation between the uncertain and nominal closed-loop behavior, the root-mean-square error (RMSE) between $\zeta^0$ and $\zeta$ is reported in Table~\ref{tab:RMS_results}. Although the visual differences in Fig.~\ref{fig:sim_result} are subtle, Table~\ref{tab:RMS_results} highlights the superior performance of realization $F^4$. Interestingly, the anticipated benefits of $F^3$ are not observed in this scenario. While the objective $\mathrm{trace}(C_z P C_z^\top)$ is designed to mitigate transient deviations in the error state $\zeta^e$, these effects appear to be negligible here. Consequently, the performance is dominated by the minimization of $\gamma$, rendering the trace objective ineffective in this specific case. Finally, $F^1$ successfully reproduces the baseline behavior of $F^0$, confirming the theoretical expectations.
\begin{table}[bt]
\centering
\caption{Comparison of Velocity RMSE and Spacing RMSE}
\label{tab:RMS_results}
\begin{tabular}{lll}
\toprule
\textbf{Realization} & \textbf{Velocity RMSE} & \textbf{Spacing RMSE} \\
\midrule
$F^0$  & $0.0238\, (-)$ & $0.1554\, (-\%)$ \\
$F^1$   & $0.0238\, (-0\%)$ & $0.1554\, (-0\%)$ \\
$F^2$   & $0.0186\, (-23\%)$ & $0.1550\, (-0.3\%)$ \\
$F^3$   & $0.0191\, (-20\%)$ & $0.1571\, (+1\%)$ \\
$F^4$   & $0.0179\, (-25\%)$ & $0.1498\, (-4\%)$ \\
\bottomrule
\end{tabular} 
\end{table}

\subsection{Experimental Results}  
\begin{figure}[bt]
    \centering
    \includegraphics[width=\linewidth]{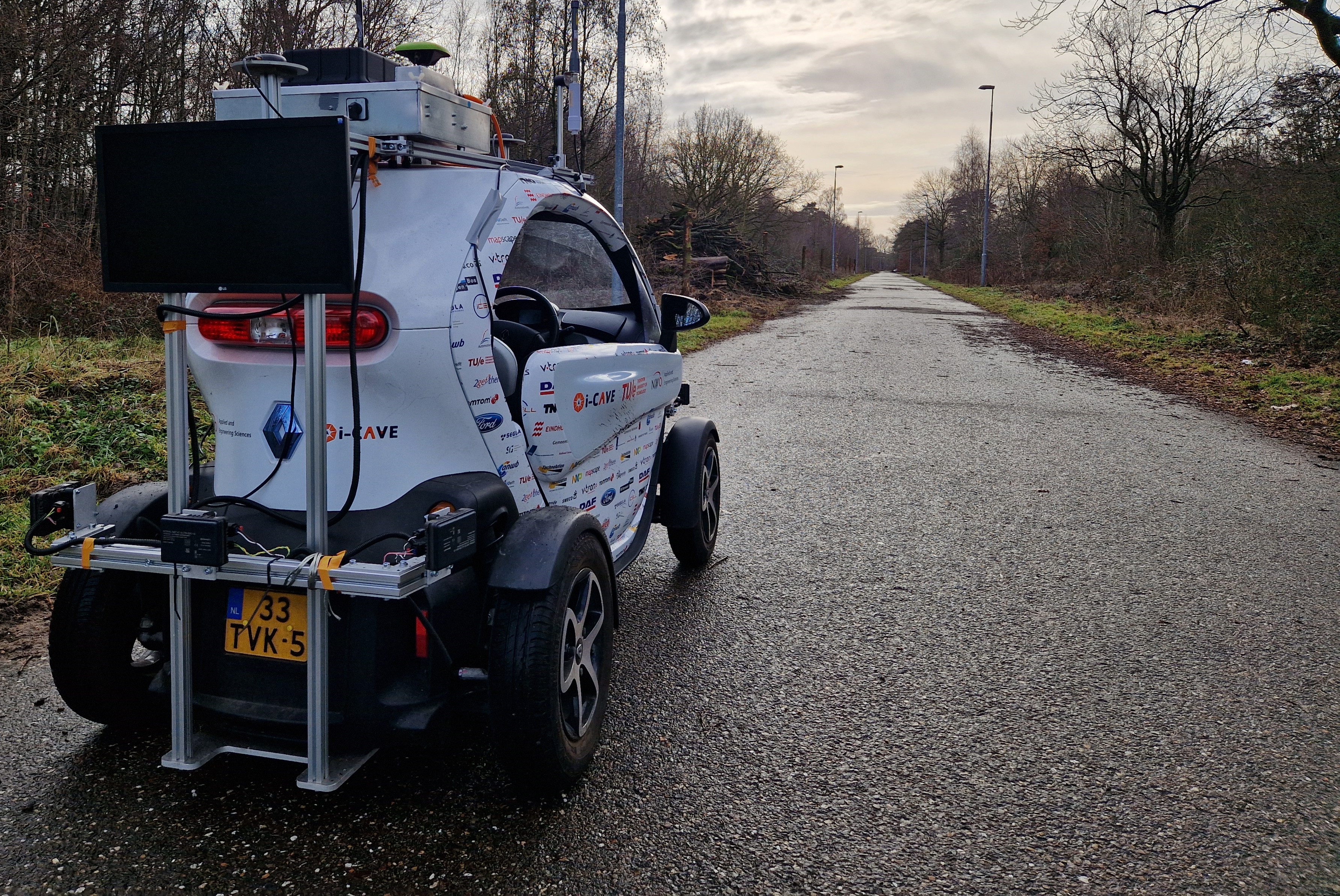}
    \caption{Renault Twizy experimental vehicle, details in~\cite{de_haan_cooperative_2025}.}
    \label{fig:twizyActionPic}
\end{figure}

\begin{figure}[bt]
    \centering
    \includegraphics[width=\linewidth]{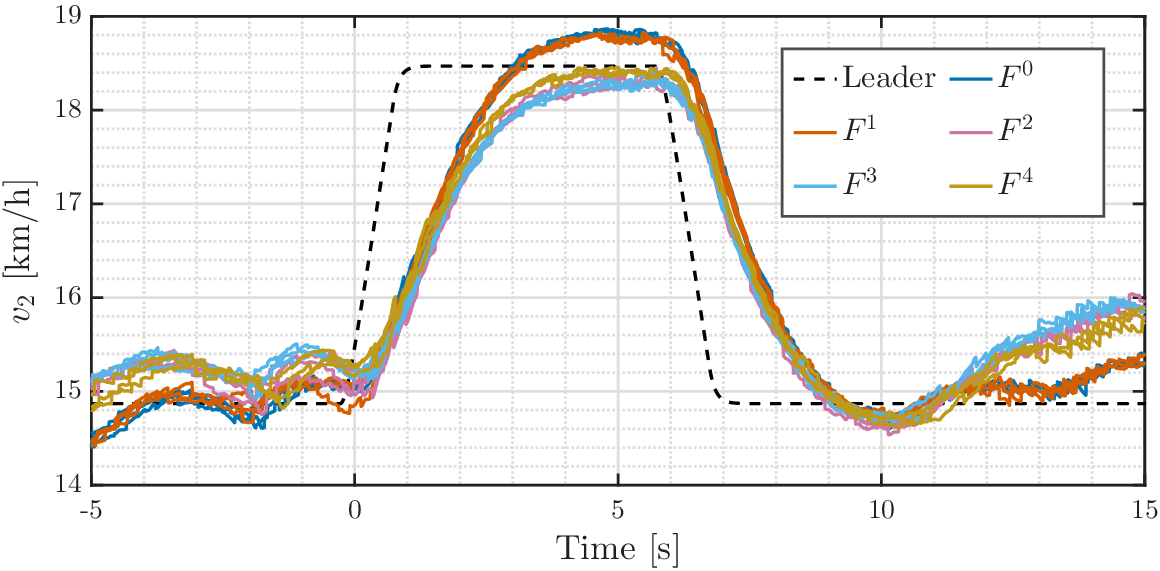} 
    \caption{Measured follower velocity $v_2$ for the PEC realizations $F^l, l\in\{1,...,4\}$, tracking a virtual leader at 15 km/h.}
    \label{fig:step_twizy}
\end{figure}

To complement the simulation study, the platooning scenario is validated experimentally using the Renault Twizy in Fig.~\ref{fig:twizyActionPic}, with details on the vehicle in~\cite{de_haan_cooperative_2025}. A virtual leader is implemented on the follower vehicle's real-time computer to enforce the desired nominal dynamics \eqref{eq:desiredModel}. The leader tracks a reference velocity $v_\text{r} = 15$~km/h and applies the same input $u_1$ as in the simulation, while the inter-vehicle distance is obtained by integrating the relative velocity. The experimental results are shown in Fig.~\ref{fig:step_twizy}. The transients observed at the start and end of the trajectory are artifacts of the cornering maneuvers required due to spatial constraints.

Overall, the experimental data align with the simulation results. Specifically, the equivalence between realizations $F^0$ and $F^1$ is confirmed, as both yield nearly identical velocity profiles. While performance differences between $F^2$, $F^3$, and $F^4$ are more subtle, realization $F^4$ consistently exhibits the best response, characterized by the smallest steady-state velocity error.

\section{CONCLUSIONS AND FUTURE WORK}\label{sec:Conclusion} 
This paper proposes a robustness-enhancing CACC framework based on PEC realizations. By exploiting alternative controller realizations that preserve the nominal control input, robustness against parametric uncertainty is improved without altering the nominal closed-loop behavior. Robust stability and performance are enforced using LMIs; experimental results demonstrate the effectiveness of the approach.

Future work will extend the framework to account for actuation delay and reformulate the LMI conditions to solve for the optimal PEC without requiring grid searches.

\printbibliography


\appendix
\subsection{Nominal System Matrices} \label{app:appendix1} \vspace{-1mm}
This appendix provides the nominal matrices for~\eqref{eq:CL_base_uncertain}.
  \begin{small}\begin{subequations}\begin{align}
    \hspace{-2mm}A^0_{i-1,i-1} \!\!&=\!\!\!
    \begin{bmatrix}
        0 & 1 \\
        0 & - \tfrac{1}{\tau\imo}
    \end{bmatrix}\!\!,\! 
    A^0_{i,i-1} \!\!=\!\!\! 
    \begin{bmatrix}
        1 & 0 \\
        0 & 0 \\
        0 & 0
    \end{bmatrix}\!\!,\! 
    B^0_{i-1} \!\!=\!\!\! \begin{bmatrix}
        0 \\ \tfrac{1}{\tau\imo}, \\
    \end{bmatrix}\!\!, \\
    \hspace{-2mm} A^0_{i,i} \!&=\! 
    \begin{bmatrix}
        0 & -1 & 0 \\
        0 & 0 & 1 \\
        0 & 0 & - \tfrac{1}{\tau_i}
    \end{bmatrix}\!, 
    B^0_i \!=\! \begin{bmatrix}
        0 \\ 0 \\ \tfrac{1}{\tau_i}, \\
    \end{bmatrix}\!,
    B^\phi_i \! = \! \begin{bmatrix}
        0 \\ 0 \\ 1 \\
    \end{bmatrix},\\
    A^c_i &=  -\tfrac{1}{h_i}, \quad
    B^c_{i,i-1} = \begin{bmatrix} \tfrac{k_d}{h_i} & \tfrac{\tau\imo - \tau_i}{h_i \tau\imo} \end{bmatrix}, \\
    B^c_{i,i}  &= \begin{bmatrix} \tfrac{k_p}{h_i} & -\left(k_p + \tfrac{k_d}{h_i}\right) & -k_d \end{bmatrix}, E^c = \tfrac{\tau_i}{h_i \tau\imo}.
\end{align}\end{subequations}\end{small}

\subsection{Uncertainty Matrices} \label{app:uncertain} \vspace{-1mm}
This appendix provides the uncertainty matrices for \eqref{eq:CL_uncertain2}:
\begin{small}\begin{subequations}
\begin{align}
    A^\Delta_{2,1} &= \begin{bmatrix} 0 & 0 & 0 \\ 0 & 0 & 0 \\ 0 & -1 & 0 \end{bmatrix}, 
    A^\Delta_{2,2} = \begin{bmatrix} 0 & 0 & 0 \\ 0 & 0 & 0 \\ 0 & 0 & -1 \end{bmatrix}, \\
    A^\Delta_{2,3} &= \begin{bmatrix} 0 & 0 & 0 \\ 0 & 0 & 0 \\ 0 & -2v_{\mathrm{r}} & 0 \end{bmatrix}, 
    A^\Delta_{2,4} = \begin{bmatrix} 0 & 0 & 0 \\ 0 & 0 & 0 \\ 0 & 0 & -v_{\mathrm{r}} \end{bmatrix}, \\
    A^\Delta_{2,6} &= \begin{bmatrix} 0 & 0 & 0 \\ 0 & 0 & 0 \\ 0 & \tfrac{p_{1|0} + 2 p_{3|0} v_{\mathrm{r}}}{p_{6|0}} & \tfrac{p_{2|0} + p_{4|0} v_{\mathrm{r}}}{p_{6|0}} - \tfrac{1}{p_{6|0} \tau_{\text{des}}} \end{bmatrix}, \\
    B^\Delta_{2,6} &= \begin{bmatrix} 0 \\ 0 \\ \tfrac{1}{p_{6|0} \tau_{\text{des}}} \end{bmatrix}, 
    B^\Delta_{0,3} = \begin{bmatrix} 0 \\ 0 \\ v_{\mathrm{r}}^2 \end{bmatrix},
    B^\Delta_{0,5} =\begin{bmatrix} 0 \\ 0 \\ -1 \end{bmatrix}, \\
    B^\Delta_{0,6} &= \begin{bmatrix} 0 & 0 & \tfrac{p_{5|0} - p_{3|0} v_{\mathrm{r}}^2}{p_{6|0}} \end{bmatrix}^\top.
\end{align}
\end{subequations}\end{small}

\end{document}